\documentclass[a4paper,11pt]{article}
\usepackage{pos}

\title{Magnetar Eruptions and Electromagnetic Fireworks}

\author*[a,b]{J. F. Mahlmann}

\affiliation[a]{Department of Astronomy \& Astrophysics, Pupin Hall, Columbia University\\ New York, NY 10027, USA}

\affiliation[b]{Department of Astrophysical Sciences, Peyton Hall, Princeton University\\ Princeton, NJ 08544, USA}

\emailAdd{jens.mahlmann@columbia.edu}

\abstract{Highly magnetized neutron stars are a source of extreme transients observed in different bands, like the fast radio burst (FRB) and associated hard X-ray burst from the Galactic magnetar SGR 1935+2154. The origin of such outbursts, hard X-rays on the one hand and millisecond duration FRBs on the other hand, is still unknown. We present a global model for various kinds of such magnetar outbursting activities. Crustal surface motions are expected to twist the inner magnetar magnetosphere by shifting the frozen-in footpoints of magnetic field lines. We discuss criteria for the development of instabilities of 3D twisted flux bundles in the force-free dipolar magnetospheres and compare their energetic properties to observations of magnetar X-ray flares. We then review a recently developed FRB generation mechanism in the outer magnetosphere of a magnetar. The strong magnetic pulse induced by a magnetar flare collides with the current sheet of the magnetar wind, compresses and fragments it into a self-similar chain of magnetic islands. Time-dependent plasma currents created during their collisions produce relatively narrow-band GHz emission with luminosities sufficient to explain bright extragalactic FRBs.}

\FullConference{High Energy Phenomena in Relativistic Outflows VIII (HEPROVIII)\\
 23-26 October, 2023\\
Paris, France\\}

\begin{document}
\maketitle

\section{Introduction}

Fast radio bursts (FRBs) are bright and short-duration ($\lesssim 1\,{\rm ms}$) radio pulses in the gigahertz band \citep[$0.1-10$\,GHz,][]{Chime2019b} with extragalactic origins \citep[see reviews by][]{Platts2019,Cordes2019,Petroff2022}. They are a new way to probe the plasma-filled universe at different scales, with fluctuations in plasma density and magnetic fields imprinted on radio wave dispersion and polarization. A growing array of radio telescopes \citep[see Section 4 in][]{Caleb2021} helped astronomers to infer some clues constraining the many theories of FRB origins. First, repeating FRBs indicated that at least some FRBs do not come from cataclysmic events \citep[e.g.,][]{Spitler2016}. Second, the simultaneous observation of FRBs and an X-ray burst from the Galactic magnetar SGR 1935+2154 show that highly magnetized neutron stars are responsible for at least some FRBs \citep{Chime2020}. 

FRBs require \emph{coherent} emission mechanisms due to their high brightness temperatures and collective propagation effects \citep[e.g.,][]{nimmo2022}. For coherent plasma emission processes, all elements in an ensemble of $N$ plasma particles interact with each other. Each interaction contributes to the coherent wave emission with a total power of $N^2$ times the emission of a single particle \citep[see, e.g.,][]{Melrose2017}. Demonstrating the validity of a coherent plasma radiation process requires first-principle modeling of plasma interactions. At this time, only two models probe the generation of FRBs and resolve the relevant plasma scales: the synchrotron maser mechanism in magnetized shocks \citep{Ghisellini2016,Plotnikov2019,Metzger2019}, and the injection of fast magnetosonic (FMS) waves by magnetic island mergers in the striped magnetar wind \citep[discussed here, and by][]{lyubarsky2020,Mahlmann2023}. X-ray bursts are likely generated by \emph{incoherent} radiation processes. For incoherent emission processes, an ensemble of $N$ plasma particles undergoes $N$ independent radiative processes with a total power of $N$ times the emission by a single particle. Roughly speaking, incoherent radiative processes are fueled by a dissipation event that provides the necessary energy, here to generate X-ray photons. The difference in radiative mechanisms is one reason why developing a theory for the simultaneous observation of X-ray bursts and FRBs is hard. In this work, we combine results from force-free electrodynamics \citep{Mahlmann2023,Rugg2023} and particle-in-cell simulations \citep{Mahlmann2022} to a global model of magnetar bursting activity. Dissipation in the inner magnetosphere incite the incoherent emission of X-rays \citep{Beloborodov2021} ranging from fainter outbursts to giant flares. Feedback of magnetic instabilities to the outer magnetosphere then stimulates the coherent generation of FRBs, either in shock-mediated or reconnection-mediated processes.

This proceedings article is structured as follows. Section~\ref{sec:globalburst} presents a global scenario for magnetar bursting activity. In Section~\ref{sec:tubeeruption}, we summarize our work on the eruption dynamics and dissipation of highly magnetized, twisted, and line-tied flux tubes \citep[][]{Rugg2023}. In Section~\ref{sec:monstershocks}, we describe how ejecta from such instabilities propagate outwards \citep[with new analyses of the work by][]{Mahlmann2023}. FMS waves can become non-linear in the extended magnetosphere and generate shocks (Section~\ref{sec:fmsout}). Large-scale eruptions expel non-linear structures that carry significant magnetic energy into the outer magnetosphere (Section~\ref{sec:bigblobs}), where they can induce FRBs (Section~\ref{sec:frbs}). We present an outlook and conclusions in Section~\ref{sec:conclusions}. 

\section{A global magnetar bursting model}
\label{sec:globalburst}

\begin{figure}
	\centering
	\includegraphics[width=0.98\linewidth]{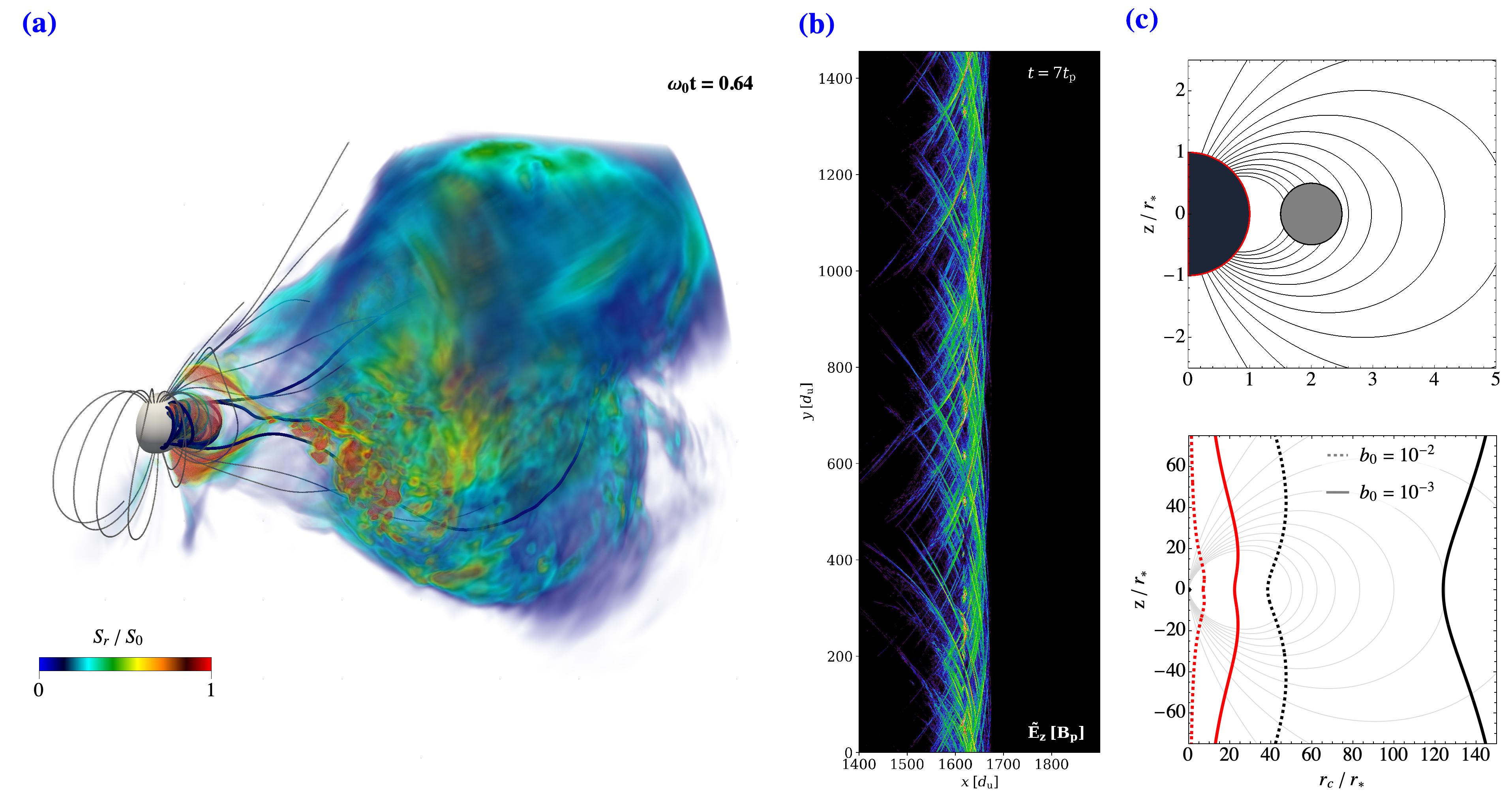}
	\vspace{-8pt}
	\caption{Different dynamical stages of the proposed magnetar bursting scenario in a global magnetospheric context. Instabilities of flux bundles twisted by crustal surface motions drive local dissipation or large-scale magnetospheric eruptions \citep[panel a, force-free electrodynamics, cf.][]{Mahlmann2023}. The energy released into the outer magnetosphere can trigger FRBs, for example by compressed reconnection of the striped magnetar wind. In this scenario, merging magnetic islands induce gigahertz radio waves \citep[panel b, particle-in-cell, cf.][]{Mahlmann2022}. FMS waves launched in the inner magnetosphere experience a significant enhancement of their amplitude $b$ relative to the background magnetic field (panel c). We indicate possible injection sites in the top panel. The injection of a spherical wave with initial amplitude $b_0$ at the stellar surface is denoted in red, injection in an extended region centered at $r_{\rm inj}=2r_\ast$ in gray color. The bottom right panel indicates where FMS waves for different injection sites become non-linear (Equation~\ref{eq:enhancementC}).}
	\label{fig:multipanel}
\end{figure}

We combine insights from several theoretical and computational works in a scenario for simultaneous X-ray and FRB emission from magnetar magnetospheres:
\begin{itemize}
    \item \emph{Magnetospheric instabilities:} Surface motions twist the magnetic field lines anchored rigidly to the magnetar crust. The stability of flux bundles with both ends tied to a conducting boundary depends on the number of field line windings per flux tube length \citep[Section~\ref{sec:tubeeruption}, see also][]{Rugg2023}. Kink modes dissipate a significant fraction of the twist energy, while higher-order instabilities can occur without much dissipation.
    \item \emph{Dissipation powers incoherent X-ray emission:} Erupting flux tubes in dipolar magnetospheres can disrupt the magnetosphere globally or dissipate energy locally \citep[Section~\ref{sec:tubeeruption}, see also][]{Mahlmann2023}. The eruption dynamics depend on the geometry of the twist-inducing surface motion. In our axisymmetric models, eruptions on global scales can dissipate giant flare-like energies of the order of $10^{46}\,{\rm erg}$. Three-dimensional eruptions on smaller scales dissipate energy in the range of typical X-ray bursts ($10^{41}-10^{43}\,{\rm erg}$). 
    \item \emph{Feedback to outer magnetosphere:} Instabilities confined to the magnetar vicinity can induce FMS waves that develop non-linear amplitudes in the extended magnetosphere. So-called electric zones can be efficient particle accelerators and likely generate shocks (Section~\ref{sec:fmsout}).
    \item \emph{Coherent FRB generation:} Energetic non-linear structures ejected during global eruptions propagate outwards \citep[Section~\ref{sec:bigblobs}, see also][]{Mahlmann2023}. The outflows transport energy to the outer magnetosphere with typical luminosities around $L_S\approx 10^{48}\,{\rm erg/s}$. Beyond the light cylinder, long wavelength pulses can generate FRBs by compressing the reconnecting current sheet of the striped magnetar wind \citep[Section~\ref{sec:frbs}, see also][]{Mahlmann2022}.
\end{itemize}

\subsection{Magnetic eruptions in the inner magnetosphere}
\label{sec:tubeeruption}

Eruptions of twisted magnetic structures in axisymmetric magnetospheres were studied with force-free electrodynamic (FFE) models in the past \citep[e.g.,][]{Parfrey2012,Parfrey2013,Mahlmann2023,Sharma2023}. FFE is the ultra-magnetized limit of ideal magnetohydrodynamics, where plasma pressure and inertial effects become negligible \citep[e.g.,][]{Mahlmann2020b,Mahlmann2020c}. The FFE limit is suitable for modeling global field line motions and identifying dissipation regions in the inner magnetar magnetosphere. In this proceedings article, we focus on the eruptions of non-axisymmetric three-dimensional (3D) flux bundles in dipolar magnetospheres \citep[][]{Carrasco:2019aas,Mahlmann2023}. The stability of line-tied flux tubes, bundles of magnetic field lines that are tied to conducting surfaces at both ends, is well-studied in the Earth and space plasma community. Rugg et al. \cite{Rugg2023} apply these insights to simplified force-free systems of straight flux tubes in a homogeneous background magnetic field. The so-called safety factor denotes the inverse of the number of field line windings per flux tube length: 
\begin{align}
    q\equiv\frac{2\pi r_0 }{L} \frac{B_z}{B_\phi} =\frac{2\pi r_0 }{Lp}\,.
    \label{eq:safetyfactor}
\end{align}
Here, we use the system length $L$, characteristic flux tube radius $r_0$, pitch factor $p=B_\phi/B_z$, and magnetic fields in cylindrical coordinates $(r,\phi,z)$. Disruption by the kink instability can occur for $q\lesssim 1$, and unstable higher order (fluting) modes can grow for $q\gtrsim 1$. Significant dissipation of $40-80\%$ of the twist energy (depending on the flux tube geometry) occurs for kink unstable systems, while higher order modes only dissipate up to $20\%$ \citep[see Figure 6 in][]{Rugg2023}. For a flux tube of a given twist, the system length regulates the instability growth: long and skinny flux tubes can have critical safety factors even for small twists. Both the fastest-growing wavelength of the kink mode, as well as its growth rate depend on the pitch factor $p$. Small values of $p$ imply a long minimum wavelength for the kink mode and small growth rates. We discuss the consequences of the pitch-dependent instability growth for dipolar magnetospheres in \cite{Rugg2023}. 

A flux tube instability can have different effects on the magnetar magnetosphere: it can dissipate energy locally and maintain the dipolar topology, or it can open up the magnetosphere on global scales with extended current sheets and ejection of non-linear magnetic structures. Mahlmann et al. \citep[][Figure~3]{Mahlmann2023} discuss a criterion for the global dynamics of flux tube eruptions that depends on how many (dipolar) field lines enclose the twisted structure. Deeply buried structures, with narrow regions of footpoint shearing induced close to the magnetic equator, are more likely to dissipate energy locally. Thicker flux tubes with footpoints closer to the magnetic pole are prone to open up the magnetosphere on large scales. While the local dissipation depends on the kind of instability (kink versus higher order modes, see above), both eruption types (localized versus global) can drive feedback on the extended or outer magnetosphere. We discuss these feedback mechanisms in the following sections.

\subsection{Feedback to the extended and outer magnetosphere}
\label{sec:monstershocks}

Crustal motions on the magnetar surface can drive instabilities of flux ropes in the inner magnetosphere. Kinking and higher-order instabilities drive the dissipation of magnetic energy in reconnection layers of different scales. This magnetic energy is then available to power the incoherent generation of abundant magnetar X-ray flares \citep{Beloborodov2021}. However, some models for X-ray bursts and FRBs rely on redistributing significant energy to the extended magnetosphere (distances up to a fraction of the light cylinder), or even the outer magnetosphere (around and beyond the light cylinder). In this section, we discuss some of these transport mechanisms.

\subsubsection{Amplitude enhancement of outwards-propagating fast magnetosonic waves}
\label{sec:fmsout}

\begin{figure}
\centering
  \includegraphics[width=1.0\linewidth]{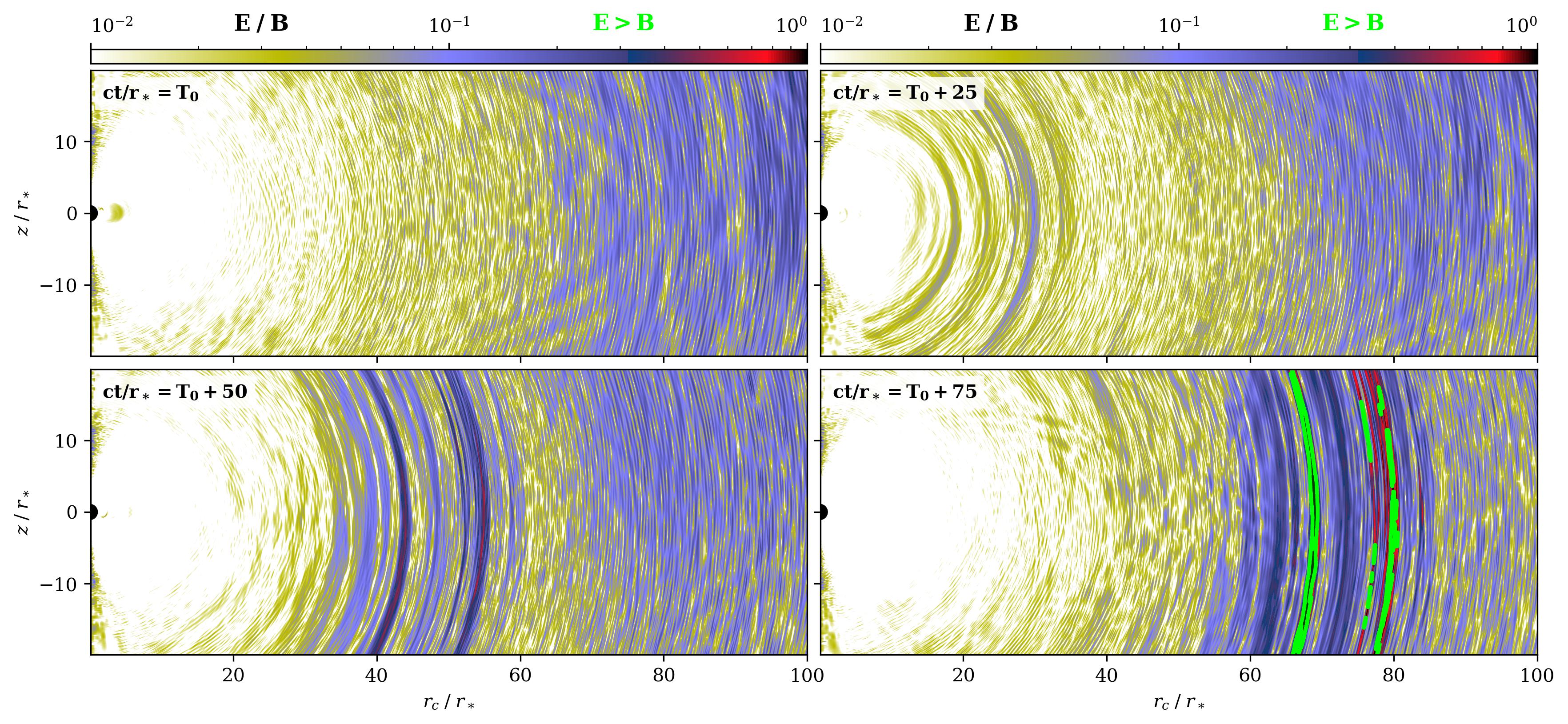}
  \vspace{-24pt}
  \caption{Electric fields of the extended magnetosphere following a confined eruption of an unstable fluxtube in the inner magnetosphere \citep[][Figure 2, left panel]{Mahlmann2023}. This figure shows 2D slices from a 3D simulation, capturing the center cross-section of a twisted fluxtube. The fluxtube instability seeds small amplitude FMS waves propagating outwards on the rapidly decaying dipole background field. At a distance of $60-80r_*$ from the injection site, the FMS waves develop electric zones with $E\approx B$ (green contours).}
  \label{fig:HOTSPOTNONLINEAR1}
\end{figure}

FMS waves injected with an amplitude $a_0$ on a source magnetic field $B_{\rm bg,s}$ propagate on a background magnetic field $B_{\rm bg}$. The initial relative amplitude $b_0=a_0/B_{\rm bg,s}$ changes due to two effects. First, the wave amplitude $a=a_0 \left(r_{\rm s}/r_{\rm fms}\right)$ decreases with distance $r_{\rm fms}$ from the source of characteristic size $r_{\rm s}$. Second, the background magnetic field $B_{\rm bg}$ can vary. FMS waves injected close to the central object will propagate mostly radially for $r\gg r_\ast$. The electric field for radial axisymmetric waves, $\mathbf{E}=E\,\mathbf{e}_\phi$, is then purely toroidal for spherical coordinates with normal basis vectors ($\mathbf{e}_r$, $\mathbf{e}_\theta$, $\mathbf{e}_\phi$). For the magnetic field $\mathbf{B}=\mathbf{B}_{\rm bg}+E\, \mathbf{e}_\theta$ we assume a dipole background field,
\begin{align}
    \mathbf{B}_{\rm bg}=\frac{2\mu\cos\theta}{r^3}\mathbf{e}_r+\frac{\mu\sin\theta}{r^3}\mathbf{e}_\theta.
\end{align}
We evaluate FMS waves that become non-linear with $E\approx B$ \citep[cf.][Equation~110]{Beloborodov2022}: 
\begin{align}
	\frac{\mathbf{E}^2}{\mathbf{B}^2}=\frac{E^2}{B_{\rm bg}^2+E^2+2 E B_{{\rm bg,}\theta}}\equiv 1\qquad\Leftrightarrow\qquad E=-\frac{B_{\rm bg}^2}{2 B_{{\rm bg,}\theta}}
	\label{eq:enhancementC}
\end{align}
This condition can be fulfilled when the wave magnetic field is anti-aligned with the background magnetic field $\mathbf{B}_{\rm bg}$, and one can reach $E=-a$. The non-linearity condition is then
\begin{align}
    2b_0=\left(\frac{B_{\rm bg}}{B_{\rm bg,s}}\right)\left(\frac{B_{\rm bg}}{B_{{\rm bg,}\theta}}\right)\left(\frac{r_{\rm fms}}{r_{\rm s}}\right)\qquad\Rightarrow\qquad \frac{r_{\rm nl}(\theta)}{r_\ast}\approx \left(\frac{4-3\sin^2\theta}{2b_0\sin\theta}\right)^{1/2}
\end{align}
Here, we estimate the radius $r_{\rm nl}(\theta)$ where FMS waves become non-linear for an injection on the stellar surface (see red lines in panel c, Figure~\ref{fig:multipanel}). For equatorial waves, the critical radius is $r_{\rm nl}/r_\ast\approx 1/\sqrt{2b_0}$. For FMS waves injected at the stellar surface, the enhancement can be significant, with $r_{\rm nl}/r_\ast\approx 70$ for $b_0=10^{-4}$. High-frequency FMS waves with wavelength $\lambda_{\rm fms} < r_\ast$ emerge ubiquitously from perturbations of the magnetar crust \citep[see appendix of][]{Yuan2022}. The simulations discussed in this work show abundant small-amplitude fast waves, for example, the background of wave-like electric fields that increase in relative amplitude for larger equatorial radii $r_c$ in Figure~\ref{fig:HOTSPOTNONLINEAR1}. Those waves, however, do not develop electric zones ($E\approx B$) for  $r_c/r_\ast\lesssim 100$. 

We evaluate the enhancement for waves injected in the inner magnetosphere in Figure~\ref{fig:multipanel} (panel c, gray circle). Magnetospheric instabilities like the confined eruption of a twisted flux rope dissipate magnetic energy locally without opening up the magnetosphere on large scales. Such disruptions induce high-frequency FMS waves of comparably large amplitudes, $b_0\approx 0.001-0.01$ (see Figure~\ref{fig:HOTSPOTNONLINEAR1}). Zones of anti-aligned wave and background fields develop electric dominance ($E\approx B$) at $r_c/r_\ast\approx 60-80$. In every time step, the FFE scheme re-establishes magnetically dominated plasma by algebraically `shaving off' any excess electric fields. In such regions (marked by green contours in Figure~\ref{fig:HOTSPOTNONLINEAR1}), the FFE integration is no longer physical, and the lack of plasma inertial effects and their feedback on the electromagnetic fields prohibits shocks and accurate dissipation rates \citep[cf.,][]{Mahlmann2020b,Mahlmann2020c}. Several authors describe the plasma reaction to such zones theoretically \citep{Beloborodov2022,Levinson2022} and in kinetic simulations \citep{Chen2022}. Particle acceleration in electric zones ($E\approx B$) can be a relevant source of plasma heating \citep{Levinson2022} or an efficient particle accelerator in shocks \citep{Beloborodov2022}. Such shocks are additional source for X-ray and FRB generation in the extended magnetar magnetosphere.

\subsubsection{Expulsion of energetic non-linear structures in large-scale eruptions}
\label{sec:bigblobs}

\begin{figure}
	\centering
	\includegraphics[width=1.0\linewidth]{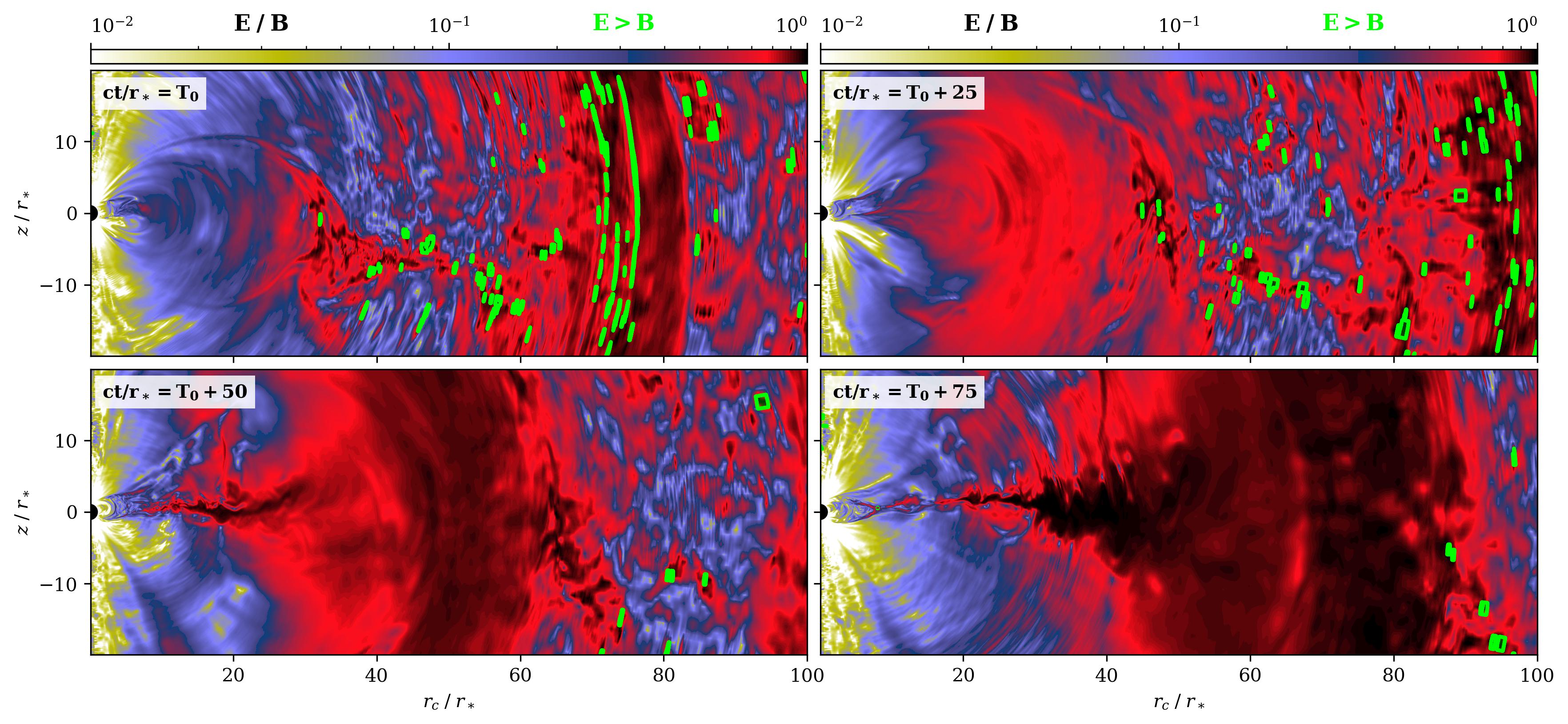}
	\vspace{-24pt}
	\caption{As Figure~\ref{fig:HOTSPOTNONLINEAR1}, but for a global disruption of the dipole magnetosphere. Initially, wave-like perturbations propagate outwards and develop electric zones ($E\approx B$). At late times, strong electric fields emerge on large scales and propagate outwards in a non-linear structure \citep[cf.][]{Sharma2023}. These structures do not necessarily become electrically dominated like their wave counterparts, and $E\lesssim B$.}
	\label{fig:HOTSPOTNONLINEAR2}
\end{figure}

Axisymmetric (2D) eruptions of the magnetar magnetosphere open up the dipole fields on large scales and develop extended current sheets \citep{Parfrey2012,Parfrey2013,Mahlmann2023,Sharma2023}. Twisted three-dimensional (3D) flux ropes with foot points close to the equator, deeply buried by the dipolar magnetic field, expel energetic structures that resemble the topology of solar coronal mass ejections \citep[CMEs, see also][]{Mahlmann2023,Sharma2023}. In their extensive simulations, \cite{Sharma2023} describe the outwards propagation of such structures in 2D, and point out the absence of extended electrically dominated zones. In 3D, \cite{Mahlmann2023} identify global eruption events for suitable flux tube geometries. Global 3D eruptions develop extended current sheets and can eject structures that carry away a relevant fraction of the twist energy induced by crustal surface motions (see Figure~\ref{fig:multipanel}, panel a). Typical luminosities of the outflowing energy `blobs' can reach $L_S\approx 5.7\times 10^{48}\,{\rm erg/s}$. In Figure~\ref{fig:HOTSPOTNONLINEAR2}, we display the extended electric field structure of such ejecta. The CME-like topology of the expelled `blob' (Figure~\ref{fig:HOTSPOTNONLINEAR2}, top right panel) propagates outwards, extends over significant distances of $\lambda_{\rm CME}/r_\ast\approx 30$. It carries strong electric fields that do not become dominant ($E\lesssim B$), different to the high-frequency FMS waves described in Section~\ref{sec:fmsout}. With their high luminosities and typical durations of 
\begin{align}
	\tau_{\rm CME}\approx 1\times\left(\frac{\lambda_{\rm CME}}{30r_\ast}\right)\left(\frac{r_\ast}{10{\rm km}}\right)\,{\rm ms}\,,
\end{align}
the outflowing non-linear structures could drive a feedback to the outer magnetosphere and generate FRBs in the striped wind around the magnetar light cylinder. 

\subsection{FRB generation by compressed reconnection}
\label{sec:frbs}

Lyubarsky \cite[][]{lyubarsky2020} and Mahlmann et al. \cite{Mahlmann2022} develop and analyze the so-called reconnection-mediated FRB generation model. Beyond the magnetar light cylinder, a long-wavelength magnetic pulse compresses the thick current sheet of a slowly rotating magnetar (kilometer layer sizes for 1s rotational periods). The reconnection layer fragments into a chain of magnetic islands (plasmoids) and, in the presence of strong synchrotron cooling, their mergers inject gigahertz-frequency radio waves (Figure~\ref{fig:multipanel}, panel b). In such plasmas, radio wave generation is a coherent emission process, and the size of merging magnetic islands determines the outgoing wave frequency. With the scaling relations predicted theoretically \cite{lyubarsky2020} and confirmed numerically \cite{Mahlmann2022}, we find that the reconnection-mediated process generates FRBs of luminosity:
\begin{align}
    L_{\rm b}=10^{42}\left(\frac{f}{2\times 10^{-3}}\right)\left(\frac{L_{\rm p}}{10^{47} \text{erg}/\text{s}}\right)^{1/2}\left(\frac{B_*}{10^{15}\text{G}}\right)\left(\frac{1\text{s}}{P}\right)\left(\frac{1\text{ms}}{\tau}\right)\text{erg}/\text{s}.
\end{align}
Here, $L_{\rm p}$ is the luminosity of the long wavelength pulse compressing the current sheet, $B_\ast$ is the stellar magnetic field, $P$ is the magnetar spin period, and $\tau$ is the duration of the burst envelope. The factor $f$ denotes the conversion efficiency between reconnected magnetic energy and injected FMS wave energy. We note that $L_S$ and $\tau_{\rm CME}$ are inferred from FFE simulations of the inner magnetosphere, and compare well to the model parameters (see Section~\ref{sec:bigblobs}). The burst luminosity and other parameters constrain the resulting wave frequency:
\begin{align}
    \nu&\approx 1\times\left(\frac{2\times 10^{-3}}{f}\right)^{5/4}\left(\frac{L_{\rm b}}{10^{42}\text{erg}/\text{s}}\right)^{5/4}\left(\frac{10^{15}\text{G}}{B_*}\right)\left(\frac{1\text{s}}{P}\right)^{3/4} \left(\frac{0.1}{\beta_{\rm rec}}\right)^{1/2}\left(\frac{100}{\xi\zeta}\right)\left(\frac{\tau}{1\text{ms}}\right)^{5/4}\text{GHz}.
    \label{eq:FRBfreq}
\end{align}
Here, $\beta_{\rm rec}$ denotes the reconnection rate, $\xi$ is the ratio between plasmoid size and current sheet thickness, and $\zeta$ measures the current sheet thickness in (relativistic) particle Larmor radii. Radio pulses produced by magnetic island mergers naturally have a downward frequency drift due to the hierarchical evolution of plasmoid sizes. We conclude that some important characteristics of FRBs could be explained by this first principle model \citep[limitations discussed in][]{Mahlmann2022}.

\section{Conclusions}
\label{sec:conclusions}

In the sequence of works discussed in this proceedings article \citep{Mahlmann2022,Mahlmann2023, Rugg2023} we lay out a scenario that could explain simultaneous X-ray and radio bursting activities from magnetar magnetospheres. We consider plasma dynamics across scales: from the disruption of flux tubes in the inner magnetosphere to their feedback on the extended magnetosphere, and kinetic processes in the striped magnetar wind. While we refer the reader to the original works for the detailed energetic scalings and limitations, this proceedings article describes the `big-picture' of our magnetar bursting model for the first time. By further evaluating simulation data from Mahlmann et al. \cite{Mahlmann2023}, we describe the feedback of magnetic instabilities close to the star on outer magnetospheric regions. FMS waves emerging from localized events can develop electric zones and drive X-ray emission from radiative shocks \citep{Beloborodov2022} or FRB injection by the synchrotron maser mechanism \citep{Ghisellini2016,Plotnikov2019,Metzger2019}. Luminous energy flows from global eruptions can become seed perturbations for so-called `far-away' FRB models, like reconnection-mediated FRB injection in the striped magnetar wind.

It is important to note that the reconnection-mediated FRB generation model has limited applicability to the observed magnetar-associated FRB from SGR1935+2154. The faint burst luminosity and gigahertz frequency cannot be recovered in Equation~(\ref{eq:FRBfreq}) for realistic parameters of this magnetar. However, the reconnection-mediated FRB generation model can be further validated and developed in (at least) three ways. First, future observations may show X-ray burst associations with more luminous FRBs. Second, reconnection can drive FMS wave injection at different locations in the magnetar magnetosphere and thereby explain the comparably low luminosities of FRBs from SGR1935+2154 \citep{yuan2020}. Third, more accurate limits on the model itself can change the frequency-luminosity dependence predicted by Equation~(\ref{eq:FRBfreq}), like different conversion factors $f$ in three-dimensional reconnection layers.

Alternatively, shock-mediated FRB generation mechanisms remain a viable candidate and require seed mechanisms. The FMS waves injected by instabilities of the inner magnetosphere as described in this proceedings article could be such a seed mechanism. Shocks driven by electric zones in such non-linear FMS waves could produce FRBs by the synchrotron maser mechanism \citep[see review in][]{Lyubarsky2021}. Determining the exact location of FRB generation requires a reliable theory of radio wave escape from the inner magnetosphere. While many of the suggested FRB models rely on the ability of radio waves to propagate through the inner magnetosphere, there are theoretical arguments against it \citep[most notably][]{Beloborodov2021a}. Besides these arguments, we still lack a first-principle confirmation of the question `Can radio waves escape from the inner magnetar magnetosphere?'. 

Modeling FRB injection and propagation across scales is a hard task, it requires capturing dynamics extending from QED-relevant scales ($\lesssim 1\,{\rm cm}$) all the way to the magnetar light cylinder (around $10^{10}\,{\rm cm}$). In this proceedings article, we adopt a global perspective that attempts to bridge across these scales by combining simulations in different regimes. This perspective is essential for the development of complete and reliable theories for the rich radiative phenomena observed from magnetars.

\section*{Acknowledgements}

The author thanks Andrei M. Beloborodov, Xinyu Li, Alexander A. Philippov, and Arno Vanthieghem for helpful discussions during the preparation of this manuscript. I acknowledge the substantial contributions of my collaborators to the works discussed in this proceedings article: Bart Ripperda, Amir Levinson, Hayk Hakobyan, Vassilios Mewes, Elias R. Most, Alexander A. Philippov, Natalie Rugg, Lorenzo Sironi, and Anatoly Spitkovsky. We are grateful for the funding provided through NASA grant 80NSSC18K1099. We acknowledge support from the National Science Foundation under grant No. AST-1909458. This research was facilitated by the Multimessenger Plasma Physics Center (MPPC), NSF grant PHY-2206607.

\bibliographystyle{JHEP}
\bibliography{literature.bib}

\end{document}